\documentclass[aip,pop,reprint]{revtex4-1}

\usepackage{graphicx}
\usepackage{dcolumn}
\usepackage{bm}
\usepackage{amsmath}
\usepackage{natbib}

\begin{document}


\title{Kinetic cascade beyond magnetohydrodynamics of solar wind turbulence in two-dimensional hybrid simulations} 



\author{D.~Verscharen}
\email{verscharen@mps.mpg.de}
\affiliation{Max-Planck-Institut f\"ur Sonnensystemforschung, Max-Planck-Str.~2, 37191 Katlenburg-Lindau, Germany}
 \affiliation{Institut f\"ur Theoretische Physik, Technische Universit\"at Braunschweig, Mendelssohnstr. 3, 38106 Braunschweig, Germany}

\author{E.~Marsch}
 \email{marsch@mps.mpg.de}
\affiliation{Max-Planck-Institut f\"ur Sonnensystemforschung, Max-Planck-Str.~2, 37191 Katlenburg-Lindau, Germany}

\author{U.~Motschmann}
 \email{u.motschmann@tu-braunschweig.de}
\affiliation{Institut f\"ur Theoretische Physik, Technische Universit\"at Braunschweig, Mendelssohnstr.~3, 38106 Braunschweig, Germany}
\affiliation{Institut f\"ur Planetenforschung, DLR, Rutherfordstr.~2, 12489 Berlin-Adlershof, Germany}

\author{J.~M\"uller}
 \email{joa.mueller@tu-bs.de}
\affiliation{Institut f\"ur Theoretische Physik, Technische Universit\"at Braunschweig, Mendelssohnstr.~3, 38106 Braunschweig, Germany}


\date{03. January 2012}

\begin{abstract}
The nature of solar wind turbulence in the dissipation range at scales much
smaller than the large MHD scales remains under debate. Here a
two-dimensional model based on the hybrid code abbreviated as A.I.K.E.F. is
presented, which treats massive ions as particles obeying the kinetic Vlasov
equation and massless electrons as a neutralizing fluid. Up to a certain
wavenumber in the MHD regime, the numerical system is initialized by assuming
a superposition of isotropic Alfv\'en waves with amplitudes that follow the
empirically confirmed spectral law of Kolmogorov. Then turbulence develops
and energy cascades into the dispersive spectral range, where also
dissipative effects occur. Under typical solar wind conditions, weak
turbulence develops as a superposition of normal modes in the kinetic regime.
Spectral analysis in the direction parallel to the background magnetic field
reveals a cascade of left-handed Alfv\'en/ion-cyclotron waves up to wave
vectors where their resonant absorption sets in, as well as a continuing
cascade of right-handed fast-mode and whistler waves. Perpendicular to the
background field, a broad turbulent spectrum is found to be built up of
fluctuations having a strong compressive component. Ion-Bernstein waves seem
to be possible normal modes in this propagation direction for lower driving
amplitudes. Also signatures of short-scale pressure-balanced structures (very
oblique slow-mode waves) are found.
\end{abstract}

\pacs{94.05.Lk, 94.05.Pt}

\maketitle 


%
%

%

\section{Introduction}

The solar wind is a dilute plasma and known to be in a highly turbulent state
\cite{tu95,horbury05}. It exhibits fluctuations in the electromagnetic field,
the plasma density, and bulk flow velocity over a wide range of
scales\cite{bruno05}. However, the nature of solar wind turbulence in the
intermediate wavenumber regime, situated between the large inertial MHD
scales and the small dissipative electron scales, is not well understood.
Especially, the role of oblique wave propagation with respect to the
background Parker field is currently under debate. Some authors favor a more
or less independent behavior of the so called slab component (parallel with
respect to the background magnetic field) and 2D turbulence (perpendicular)
\cite{montgomery81,bieber96,oughton98}. While in this picture the highly
oblique 2D-turbulence is believed to be an example of a strongly turbulent
plasma state with high-order correlations between the fluctuating quantities,
the slab component is assumed to be describable within the framework of weak
turbulence theory, i.e.~as a superposition of normal modes. Other authors
interpret the observed anisotropy as being the result of a kinetic Alfv\'en
wave (KAW) cascade \cite{howes08,schekochihin08}, which would suggest a
preference of fluctuations with $k_{\perp}\gg k_{\parallel}$.

Incompressible, dissipative MHD simulations in two dimensions
\cite{shebalin83} and three dimensions \cite{oughton94} show the development
of an anisotropy in turbulent spectra with respect to the background magnetic
field. In these cases, an initially isotropic spectrum is shown to evolve
towards an anisotropic configuration with preferred wavevectors perpendicular
to the background field. These authors have shown that the strength of
dissipation in the simulations is crucial for the development of anisotropy.
The theoretical basis of the geometry in the presence of a strong background
field was discussed by Montgomery and Turner \cite{montgomery81} in the
framework of a perturbation analysis of the incompressible MHD equations. The
resonant conditions $\vec k_1+\vec k_2=\vec k_3$ for the wavevectors and
$\omega_1+\omega_2=\omega_3$ for the wave frequencies in a resonant
three-wave interaction were found to favor a perpendicular cascade and lead
to a suppression of energy transfer in the parallel direction. The
strongest turbulent coupling is found to occur if all three frequencies are
about zero, a situation which is closest to the hydrodynamic case. This can
be completely achieved only if the three-dimensional nature of the solar wind
fluctuations is fully resolved, like with a 3-D code that permits to consider
both dimensions perpendicular to the background field. However, it was
claimed that higher-order couplings could in fact also yield a parallel
cascade, yet with a slower growth. These higher-order couplings, the
dispersion of the normal modes, and possible compressive effects lead to
additional complications in the kinetic regime and make a numerical treatment
necessary.

Measurements made by the four Cluster spacecraft have provided further
insights into the nature of solar wind turbulence, since multi-spacecraft
detections of magnetic fluctuations even permit the analysis of their
three-dimensional dispersion properties. However, these measurements support
different interpretations. A fully evolved nonlinear turbulent state with a
preferred 2D-component beyond the MHD range was found by Alexandrova et
al.\cite{alexandrova08} The dispersion relation of this mainly
perpendicularly structured component might reflect a nonlinear cascade that
also occurs at small scales. However, an interpretation on the basis of
normal modes, such as the right-handed circularly polarized fast/whistler
(F/W) waves, is still possible \cite{narita11}.

There is strong evidence that the field-parallel component consists at least
partly of Alfv\'en/ion-cyclotron (A/IC) waves, which are dispersive left-hand
circularly polarized electromagnetic normal modes of a plasma. The
temperature anisotropies and beam structures observed in the solar wind
proton distribution functions were explained as resulting from
cyclotron-resonant interactions of the ions with these waves
\cite{marsch04,heuer07,bourouaine10}. Recently, direct wave measurements have
confirmed the existence of A/IC waves in the solar wind \cite{jian09}. Yet,
due to cyclotron-resonant wave-particle interactions, they are strongly
damped at wavenumbers corresponding to the inverse inertial length of the
resonant ions \cite{ofman05}, and thus they are not expected to exist at
higher wavenumbers. A coexistence of left-handed and right-handed modes has
been recently supported by measurements of the angle distribution of the
magnetic helicity in the solar wind \cite{he11a}. Possible candidates for
normal modes beyond the resonant wavenumber range are the F/W modes, which
may remain after the dissipation of the A/IC waves \cite{stawicki01}, or an
ongoing cascade \cite{howes08} of dispersive KAWs. Both these wave modes are
right-hand polarized under the conditions prevailing in the solar wind. It is
observed that the right-hand polarized waves survive the spectral break,
which indicates the transition from the inertial range to the ion dissipative
scales, and that they can exist at higher wavenumbers without damping, until
the resonant electron scales are finally reached \cite{goldstein94}.

There are other numerical hybrid models available, which treat the kinetic
cascade of turbulence beyond the MHD scales. For example Parashar et
al.\cite{parashar09} achieve an efficient turbulence driving by assuming an
Orszag-Tang vortex in the plane perpendicular to the background magnetic
field. This configuration leads to perpendicular ion heating, which seems to
be not connected to cyclotron-resonant effects. In a later
work\cite{parashar10}, the authors also included electron pressure effects
and found indications for Bernstein waves. However, the spectrum did not show
significant presence of Alfv\'en or whistler waves but rather a dominance of
zero-frequency structures for their geometry and in the strongly turbulent
regime. These simulations show that the presence of the electron pressure
term in the generalized Ohm's law can substantially modify the properties of
the modeled turbulence. A similar situation has recently been treated in
another work\cite{markovskii11} with a special focus on the dissipation of
the energy. The formation of intermittent current sheets on short time scales
and the related demagnetization of the plasma ions lead to a stochastic
heating \cite{chandran10}.

Our numerical simulation work is focused on studying the transition of
isotropic MHD turbulence to non-isotropic kinetic fluctuations on
intermediate scales. For this purpose, at least two-dimensional numerical
simulations are necessary, by which one can analyze the evolution of
turbulence in different directions with respect to the constant background
magnetic field. The model system is initialized by a superposition of linear
MHD waves. No further external or ongoing driving force is applied, and thus
the system will evolve freely from its initial state. Two separate simulation
runs are presented, which only differ in the amplitude of the initial
perturbation.

\section{Numerical method}

The A.I.K.E.F. code is a numerical hybrid code, which treats ions as
particles following the characteristics of the Vlasov equation and electrons
as a massless charge-neutralizing fluid. It is based on the work by Bagdonat
and Motschmann\cite{bagdonat02}, and its basic properties have later been
described extensively by Bagdonat \cite{bagdonat05}. The code was completely
revised, and the possibility for adaptive mesh refinement was included by
M\"uller et al. \cite{mueller11} The equations of motion for a single proton
are
\begin{align}
\frac{\mathrm d\vec v_{\mathrm p}}{\mathrm d t}&=\frac{q_{\mathrm p}}{m_{\mathrm p}}\left(\vec E+\frac{1}{c}\vec v_{\mathrm p}\times \vec B\right), \label{eqmotion_fla}\\
\frac{\mathrm d\vec x_{\mathrm p}}{\mathrm dt}&=\vec v_{\mathrm p}
\end{align}
for the velocity $\vec v_{\mathrm p}$ and the spatial location $\vec
x_{\mathrm p}$. The force acting on the particle with charge $q_{\mathrm p}$
and mass $m_{\mathrm p}$ is the Lorentz force, which is due to the electric
field $\vec E$ and the magnetic field $\vec B$. The speed of light is denoted
by $c$. The electron fluid equation of motion delivers the electric field as
\begin{equation}
\vec E=-\frac{1}{c}\vec u_{\mathrm e}\times \vec B-\frac{1}{n_{\mathrm e}e}\nabla p_{\mathrm e},
\end{equation}
where the electron bulk velocity is denoted by $\vec u_{\mathrm e}$, the
electron number density by $n_{\mathrm e}$, the elemental charge by $e$, and
the electron fluid pressure by $p_{\mathrm e}$. The electrons are assumed to
be isothermal, and thus the pressure depends on the electron number density
according to $p_{\mathrm e}\propto n_{\mathrm e}$. Quasi-neutrality requires
that $n_{\mathrm p}=n_{\mathrm e}$. The proton density and the proton bulk
velocity $\vec u_{\mathrm p}$ are obtained as the first two moments of the
proton distribution function at every step in time.

The magnetic field is obtained from the induction equation following from
Faraday's law:
\begin{equation}
\frac{\partial \vec B}{\partial t}=\nabla \times \left(\vec u_{\mathrm p}\times \vec B\right)-\nabla \times \left(\frac{c}{4\pi \varrho_{\mathrm c}}\nabla \times \vec B\times \vec B\right).
\end{equation}
The ion charge density $\varrho_{\mathrm c}$ corresponds to the proton number
density calculated as the zeroth velocity moment of the distribution function
of all protons.

The boundaries of the simulation box are set to be periodic, and the
particles are initialized with a Maxwellian velocity distribution that is
shifted to the given values for the initial bulk velocities. The width of the
Maxwellian distribution is determined by the respective species' plasma beta,
which represents the ratio of the ion thermal to the magnetic energy density.
The beta is set to $\beta_{\mathrm p}=0.05$ for the protons. The electron
beta is fixed at $\beta_{\mathrm e}=0.5$. In this beta regime, even low
amplitudes of magnetic fluctuations will have a strong influence on the
motion of the particles due to their high magnetization. All spatial length
scales are normalized and given in units of the proton inertial length,
$\ell_{\mathrm p}=c/\omega_{\mathrm p}$, with the proton plasma frequency
$\omega_{\mathrm p}=\sqrt{4\pi n_{\mathrm p}q_{\mathrm p}^2/m_{\mathrm p}}$.
All time scales are in units of the inverse proton gyro-frequency
$\Omega_{\mathrm p}=q_{\mathrm p}B/(m_{\mathrm p}c)$. In these units, the
two-dimensional integration box has a size of $250\times 250$, which is
covered by $1024\times 1024$ cells, each of which filled with 500
superparticles representing the real number density of the protons. In this
spatial configuration, one grid step corresponds to about $0.24\ell_{\mathrm
p}$ in the calculation. The data is recorded with a grid resolution of
$\Delta x=0.5\ell_{\mathrm p}$ to keep the amount of data reasonable. This
corresponds to a maximum resolvable wavenumber of $\ell_{\mathrm
p}k_{\mathrm{res}}\approx \pi /0.5\approx 6.3$, which is adequate for an
appropriate coverage of the transition from MHD into the kinetic regime. For
all vector quantities, three components are evaluated. Approaches of this
form are sometimes referred to as 2.5D simulations. A divergence-cleaning
algorithm is applied to guarantee numerical stability.

The initial magnetic field is given as a superposition of linear Alfv\'en
waves according to
\begin{equation}
\delta  B_x=\sum \limits _{m=1}^{m_{\max}} \sum \limits_{n=1}^{n_{\max}} b_n \cos(k_{n}y\sin\vartheta_m+k_{n}z\cos\vartheta_m+p_{n,m} ),
\end{equation}
where the constant background field $\vec B_0$ is aligned along the $z$-axis.
A random phase shift $p_{n,m}$ is applied to each wave. The amplitudes $b_n$
are fixed in such a way that the power spectrum follows a Kolmogorov power
law\cite{kolmogorov41} in wavenumber with the scaling $\propto k^{-5/3}$. The
total power of the composed wave field is made equal to the power of a
monochromatic wave with $\delta B=0.01 B_0$ for Run A and to $\delta B=0.1
B_0$ for Run B. The first value corresponds to a situation with a weaker
turbulence level or a higher background field as prevailing in the solar
corona, while the second value represents realistic solar wind conditions.
The angles $\vartheta_m$ cover $360^{\circ}$ of propagation directions by
$m_{\max}=50$ discrete values, and the spectrum ranging between $k_0=0.05$
and $k_{\max}=0.2$ is covered by $n_{\max}=20$ waves. The upper limit
$k_{\max}$ is quite high compared to typical MHD scales but, still, in the
dispersion-less range in first order. The initial velocity is obtained from
the Alfv\'enic polarization relation
\begin{equation}
\frac{\delta \vec V}{V_{\mathrm A}}=\mp \frac{\delta \vec B}{B_0}
\end{equation}
with the Alfv\'en speed $V_{\mathrm A}=\ell_{\mathrm p}/\Omega_{\mathrm p}$.

\section{Results}

In this section some results of the numerical simulation runs are shown. We
ran the code for a sufficiently long time so that an evolved nonlinear
dynamic plasma state could be expected. This was typically the case after an
evolution time of about 500 gyro-periods, at which time the system was
analyzed.

\subsection{Results for simulation Run A}

\begin{figure}
\includegraphics[width=\columnwidth]{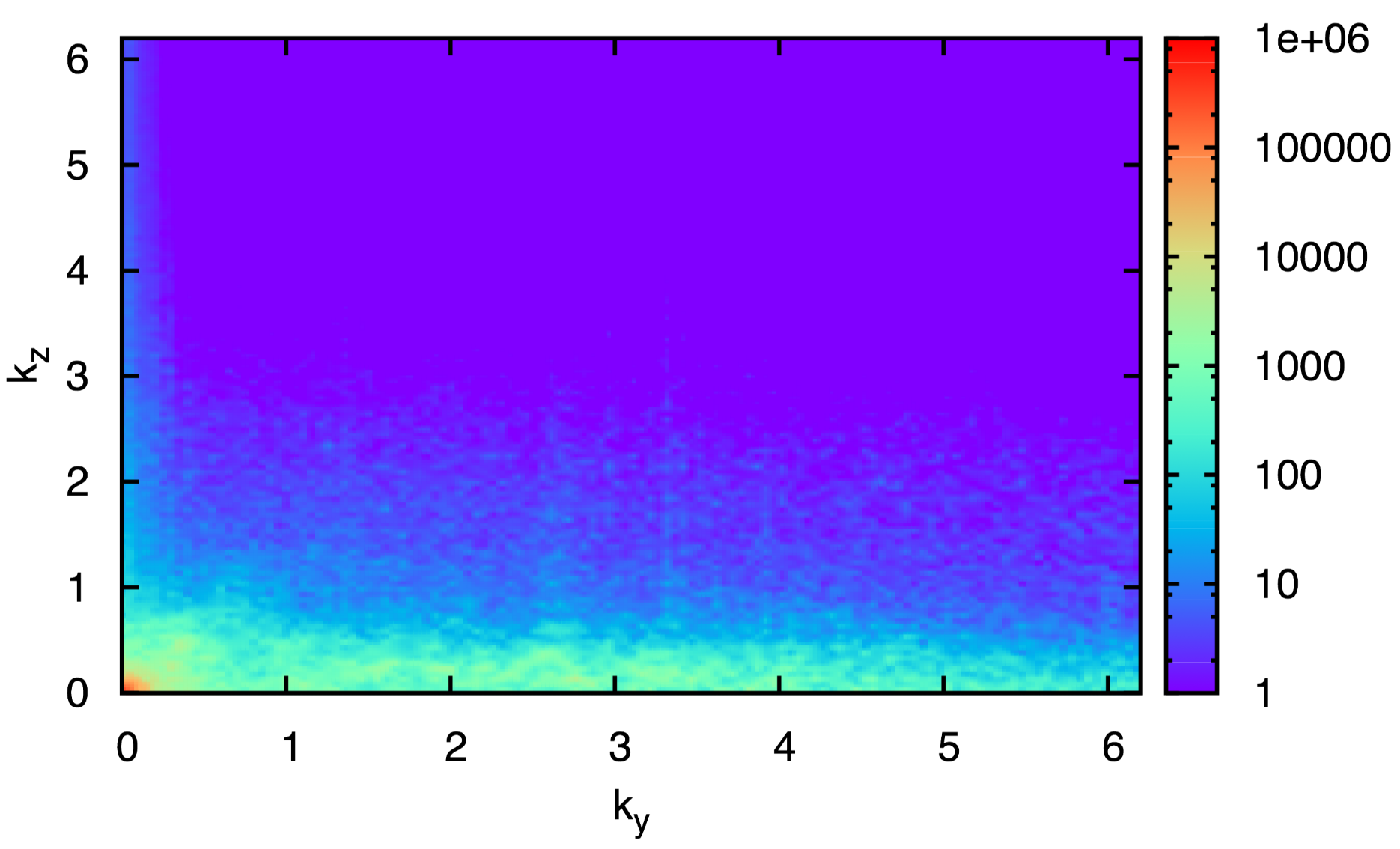}%
\caption{Two-dimensional power spectral density of magnetic field fluctuations
in arbitrary units for Run A. The background magnetic field is oriented along the $k_z$-axis.
The cascade of energy to higher wavenumbers occurs preferentially in
perpendicular direction.
\label{fig_bmhd_B_spectrum_later}}%
\end{figure}

\begin{figure}
\includegraphics[width=\columnwidth]{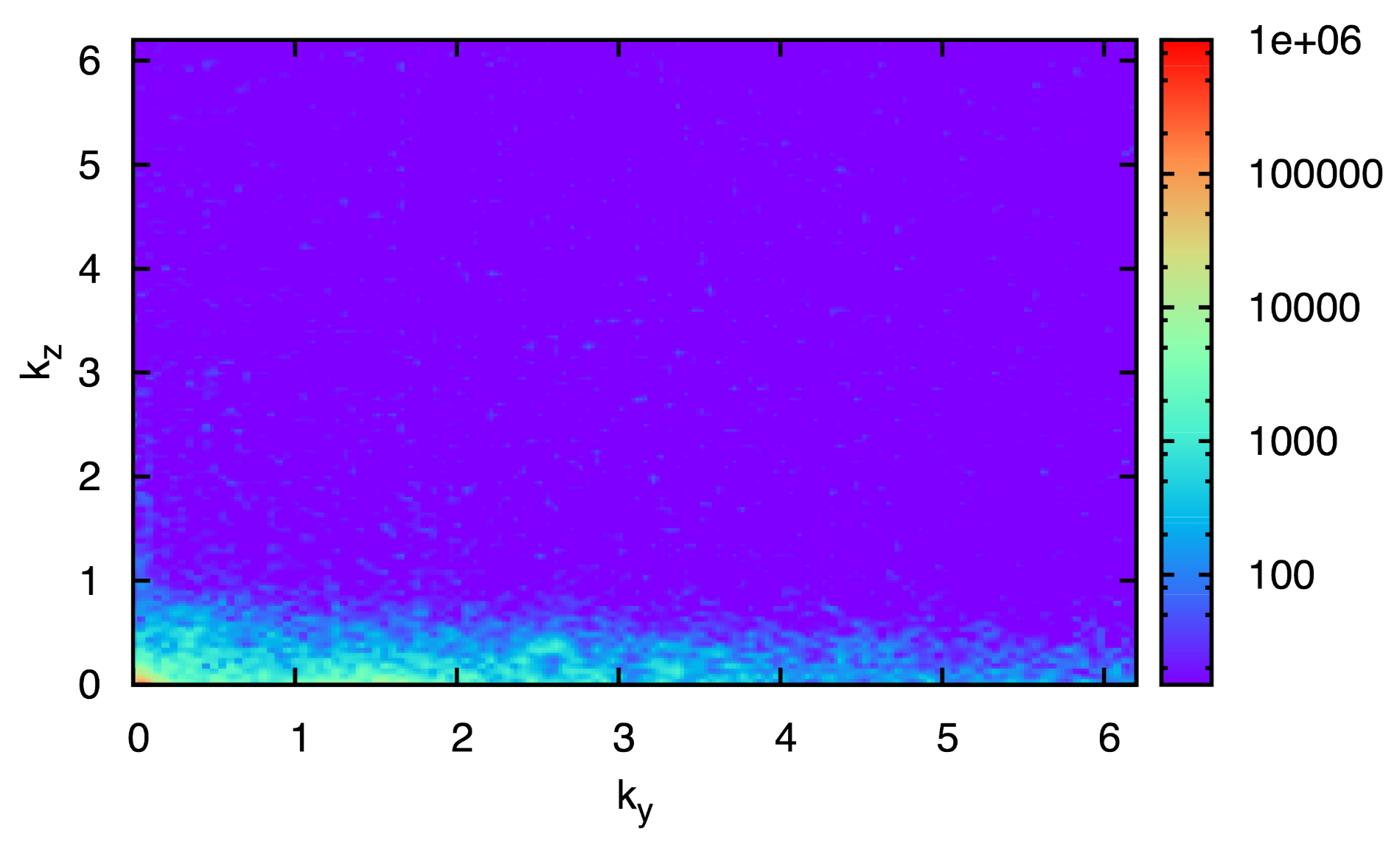}%
\caption{Power spectral density of compressive fluctuations in the proton
number density in arbitrary units for Run A. The spatial variation of the fluctuations
occurs mainly perpendicular to the background magnetic field.
\label{fig_bmhd_rho_spectrum}}%
\end{figure}

For the analysis of the results from Run A, Fourier transformations in two
dimensions were applied subsequently to the magnetic field data and the
density data. The power spectral density was then calculated, and could be
shown to depend on the wavenumbers $k_z$ for the direction parallel to the
background field and $k_y$ perpendicular to it. The resulting magnetic field
power spectral density is shown in Fig.~\ref{fig_bmhd_B_spectrum_later}, and
the power spectral density of the compressive fluctuations is shown in
Fig.~\ref{fig_bmhd_rho_spectrum}.

Apparently, the magnetic field fluctuations show a preferred alignment with
the direction perpendicular to the background magnetic field. Especially,
turbulence energy is spreading in this direction to much higher wavenumbers
than in the parallel direction. But also the parallel fluctuations at
wavenumbers beyond the initialized range are excited and gain energy. The
compressive density fluctuations, however, are mainly aligned perpendicular
to the background field and have almost no components parallel to $\vec B_0$.

\begin{figure}
\includegraphics[width=\columnwidth]{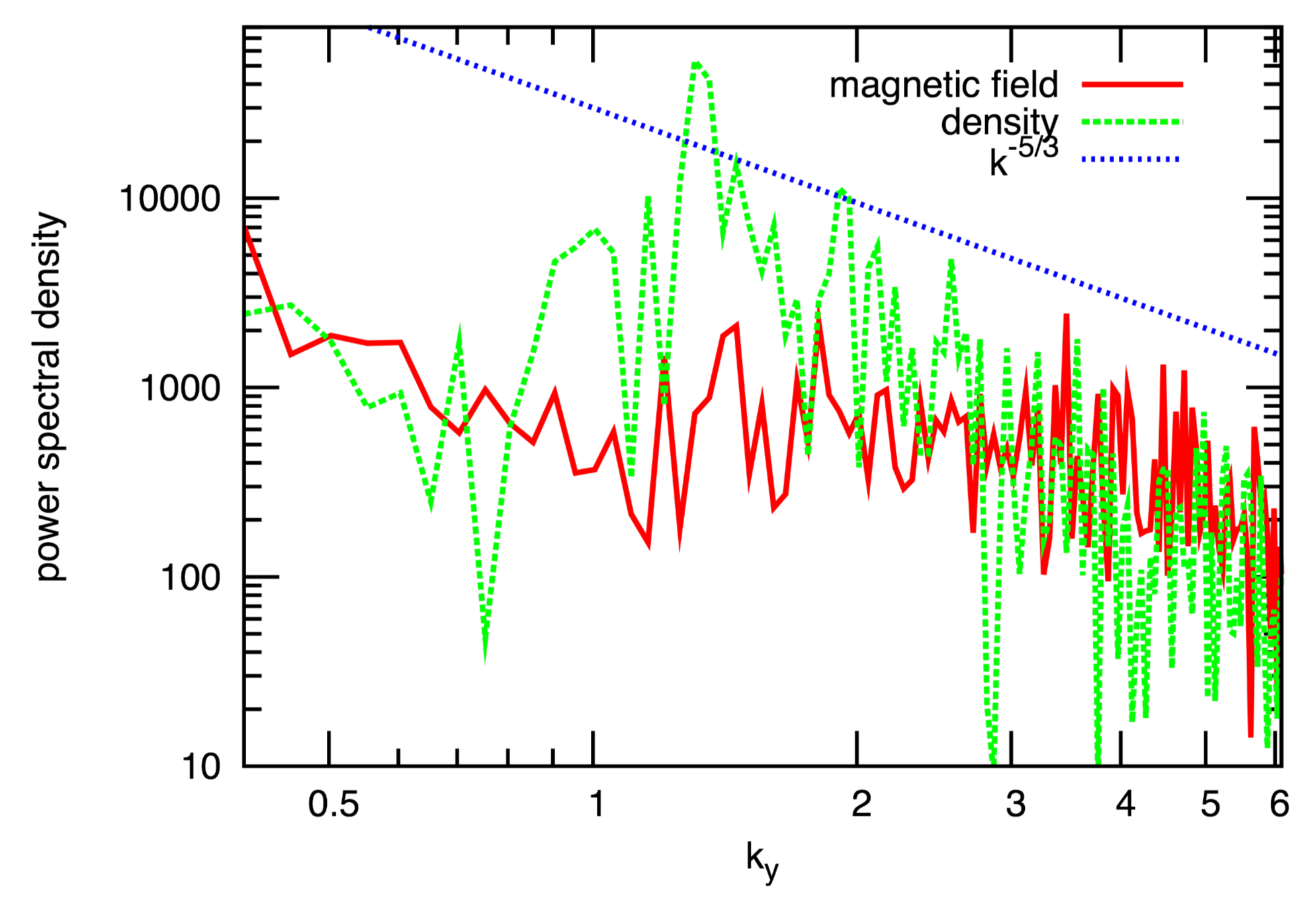}%
\caption{Power spectral density of fluctuations in the magnetic field and
density at $t=500$ along the perpendicular direction for Run A.
A power law with spectral index $-5/3$ is shown additionally, in order to
estimate the slope of the spectrum.
\label{fig_bmhd_cutspec}}%
\end{figure}

Cuts through the two-dimensional spectra along the perpendicular direction
are shown in Fig.~\ref{fig_bmhd_cutspec}.
The power spectral density of the compressive fluctuations in the proton
number density is enhanced for $k_y\gtrsim 1$ and follows mainly a power law
with a slightly steeper index than $k^{-5/3}$ in this range. The magnetic
field spectrum is flatter in the dispersive range.

\begin{figure}
\includegraphics[width=\columnwidth]{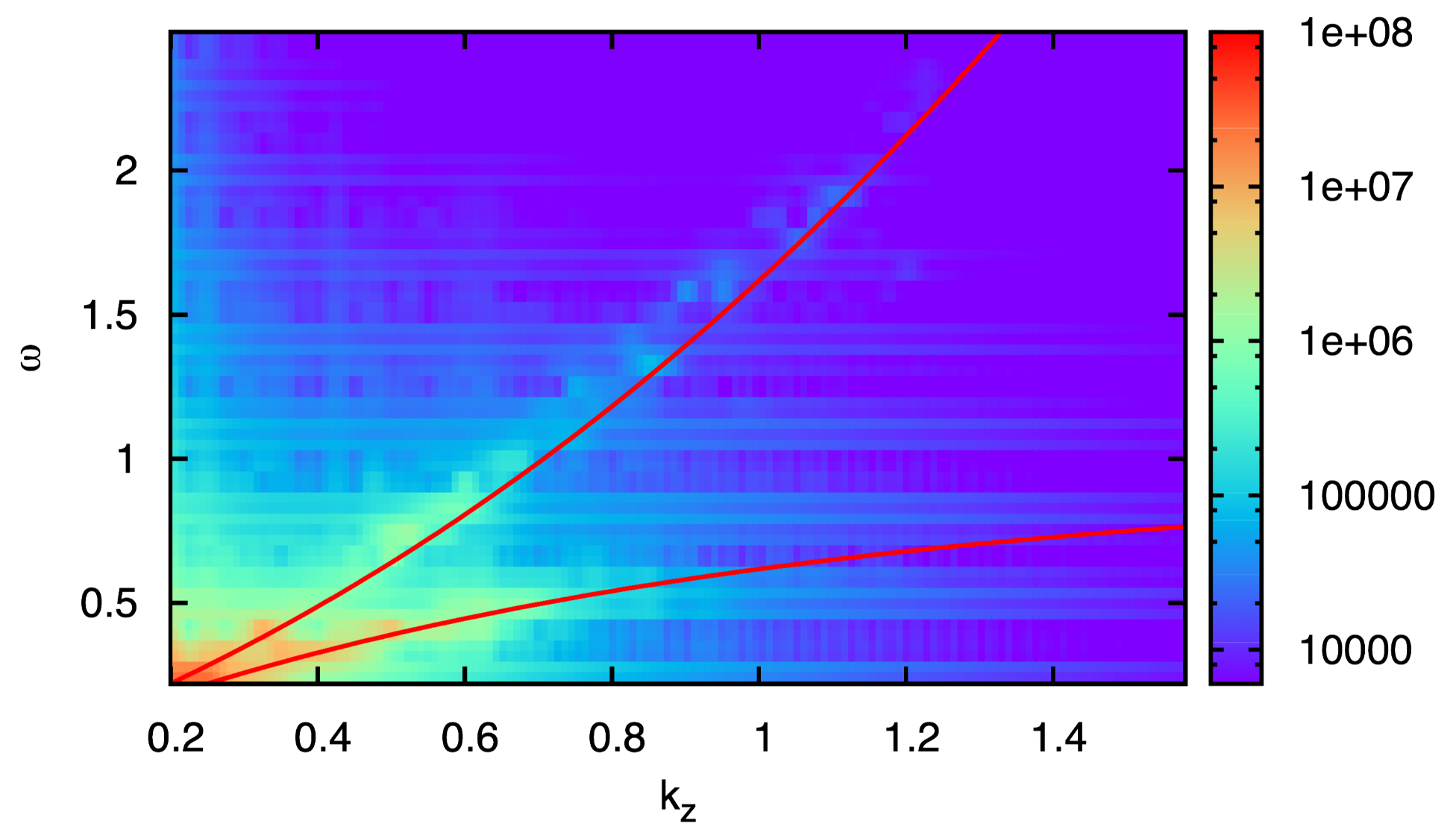}%
\caption{Dispersion relation of magnetic field fluctuations parallel to the
background magnetic field at $t\approx 500$ for Run A. The overplotted red lines
indicate the F/W (upper line) and A/IC (lower line) wave branches calculated
directly from the cold-plasma dispersion relation.
\label{fig_bmhd_dispersion_B_parallel}}%
\end{figure}

To study the nature of these fluctuations Fourier transformation can also be
used, but here is applied in the time domain leading to the corresponding
dispersion diagrams. Therefore, first the two-dimensional spatial Fourier
transformation is applied, and the result is taken in one dimension only
(parallel or perpendicular to $\vec B_0$), and then the data are again
Fourier transformed yet in time. The result for the parallel magnetic field
dispersion is shown in Fig.~\ref{fig_bmhd_dispersion_B_parallel}.

Most of the power is found at low wavenumbers and low frequencies and
still close to the initial distribution of the waves. Spreading of power at
low parallel wavenumbers with low frequencies is an indication for ongoing
MHD turbulence. Two sharp branches can clearly be seen beyond the initial
wavenumber limit at $k_{\max}=0.2$. For an easier identification of them, the
cold-plasma dispersions for the left-handed A/IC waves and for the
right-handed F/W waves are additionally shown in the parallel dispersion
diagram\cite{stix92}. The observed parallel dispersion agrees well with that
of the linear normal modes. The A/IC branch ends at a wavenumber value below
$1$, whereas the right-handed branch continues to higher wavenumbers and
frequencies.

\begin{figure}
\includegraphics[width=\columnwidth]{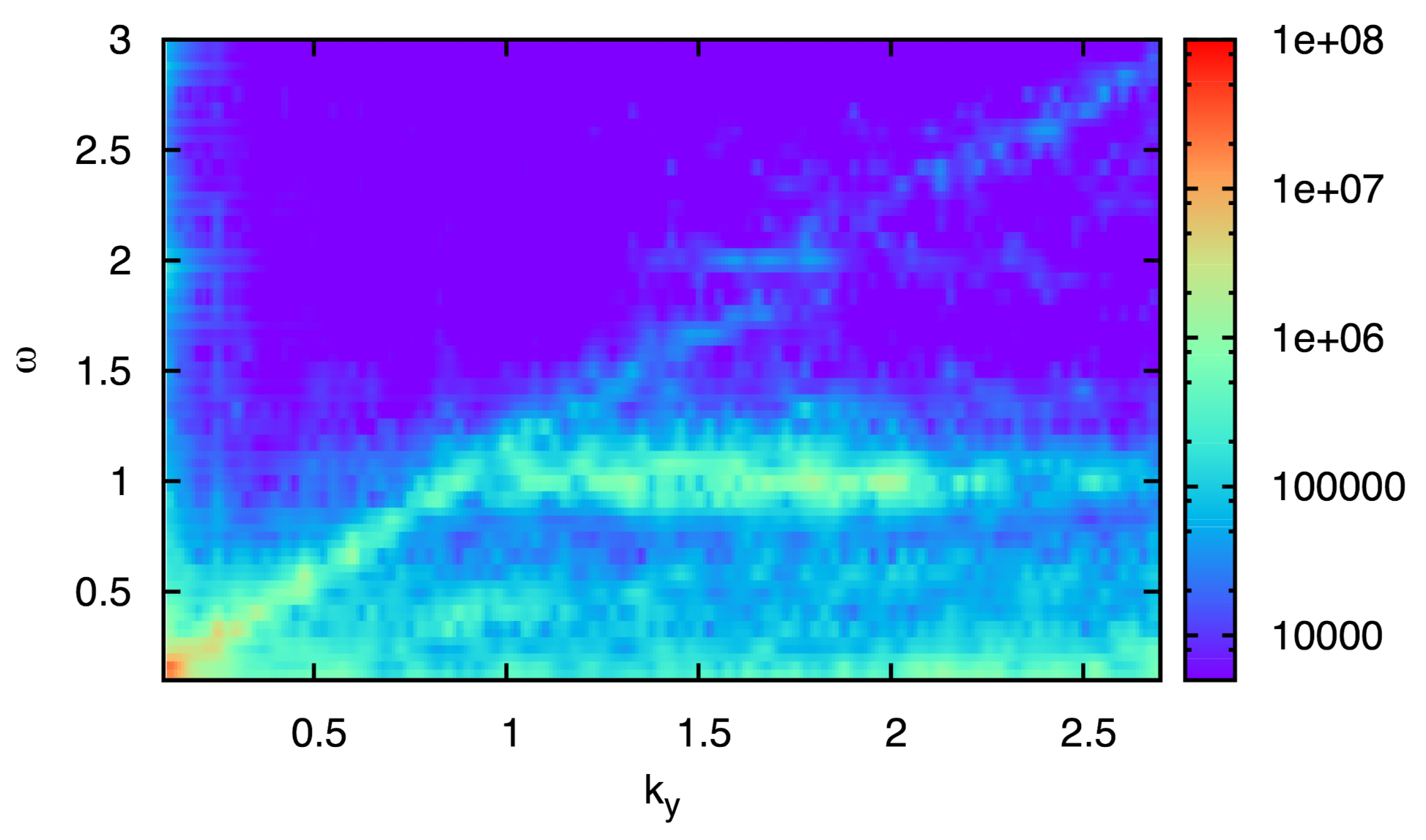}%
\caption{Dispersion relation of magnetic field fluctuations perpendicular to
the background magnetic field at $t\approx 500$ for Run A.
\label{fig_bmhd_dispersion_B_perp}}%
\end{figure}

\begin{figure}
\includegraphics[width=\columnwidth]{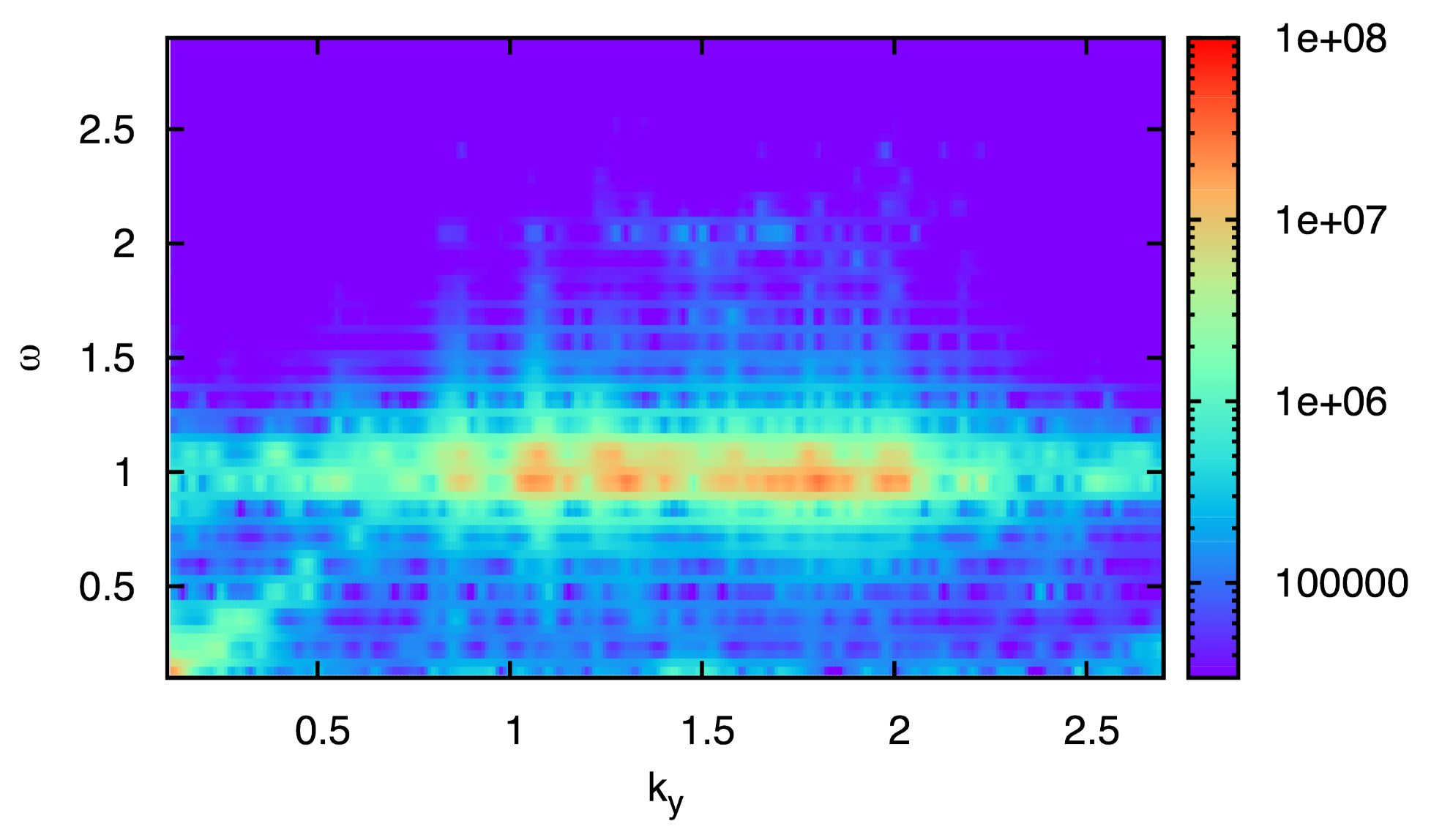}%
\caption{Dispersion relation of density fluctuations perpendicular to the
background magnetic field at $t\approx 500$ for Run A. The fluctuations
coincide well with the magnetic fluctuations.
\label{fig_bmhd_dispersion_rho_perp}}%
\end{figure}

The perpendicular magnetic field dispersion is depicted in
Fig.~\ref{fig_bmhd_dispersion_B_perp}. Since the density fluctuations are
mainly perpendicularly oriented, their parallel dispersion diagram is not
shown here. But the perpendicular dispersion diagram is illustrated in
Fig.~\ref{fig_bmhd_dispersion_rho_perp}.

These two dispersion diagrams show a common structure in the fluctuations in
the form of a band signature of the intensity near the gyro-frequency and at
higher harmonics in the magnetic and compressive dispersion. We believe that
this is a typical indication for the occurrence of ion-Bernstein
waves\cite{stix92,brambilla98,swanson03}. A linear branch with weak power is
observed at $\omega/k=1$ in the perpendicular dispersion plots of the
magnetic field and density fluctuations. It corresponds to the linear
fast-mode wave in a low-beta plasma. A linear Alfv\'en wave does not
propagate perpendicular to the background magnetic field, and therefore can
be excluded from the interpretation of this branch. A coupling between it and
the ion-Bernstein modes is not observed, and may presumably not be detectable
given the spectral resolution of the numerical analysis applied.

There is further power distributed in compressive and magnetic structures at
$\omega\approx 0$. These signatures cannot be explained as ion-Bernstein
waves, because they do not have the minimum frequency of $\Omega_{\mathrm p}$
and are merely spatial structures constant over time. To analyze their
nature, the possible correlation between the magnetic pressure $P_{B}=\delta
\vec B^2/(8\pi)$ and the density fluctuations $\delta n$ can be applied. It
is defined as
\begin{equation}
C\equiv\frac{\langle\delta n \,\delta |\vec B|^2 \rangle}{\sqrt{\langle\delta n^2\rangle\langle\delta |\vec B|^4 \rangle}},
\end{equation}
where the brackets indicate a certain way of averaging. This averaging is
here done only perpendicular to the magnetic field and in the time domain.
Averaging over a long time scale corresponds to structures with low frequency
and averaging over short time scales corresponds to higher frequencies. The
same is true in the spatial domain.

\begin{figure}
\includegraphics[width=\columnwidth]{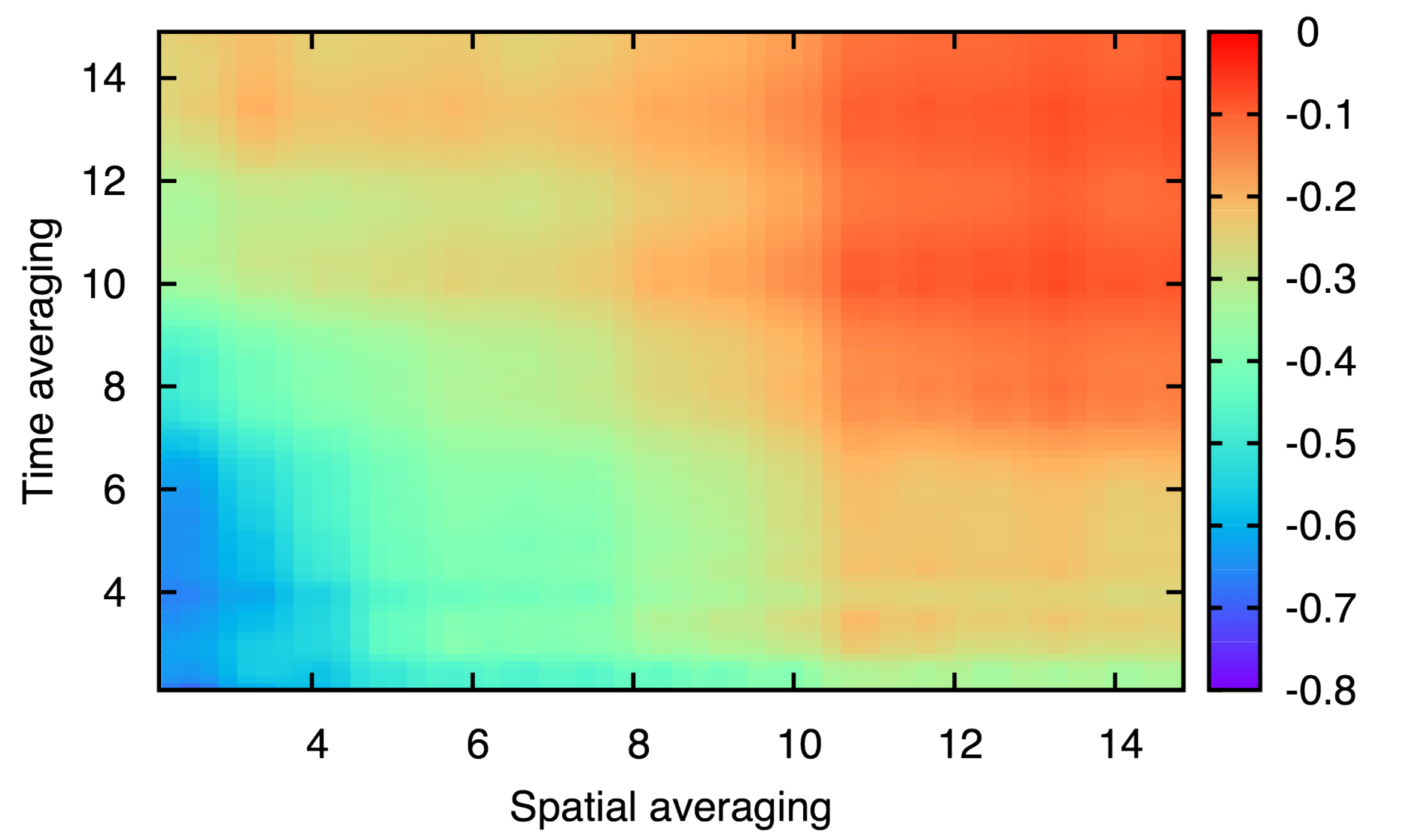}%
\caption{Correlation coefficient $C$ between fluctuations of magnetic pressure and density
depending on the averaging in space and time for Run A. A strong anti-correlation is found up to long averaging times, which correspond to low frequencies in the dispersion diagrams. This is an indication for the existence of PBSs. The units are given in the above introduced normalization.
\label{fig_bmhd_correlation}}%
\end{figure}

The result of this calculation is shown in Fig.~\ref{fig_bmhd_correlation}.
At high values for the time averaging and at low values for the spatial
averaging, a strong anti-correlation between $\delta P_B$ and $\delta n$ is
found. This indicates the existence of pressure-balanced structures (PBSs),
which correspond to steepened-up slow-mode waves in the perpendicular
direction \cite{tu94}. They are characterized by their direction of
propagation and the strong anti-correlation between $\delta P_B$ and $\delta
n$. The classical slow-mode wave does not propagate at 90 degrees, but in
this direction transforms into a tangential discontinuity, which some of the
numerical structures may represent. Therefore, it seems useful to analyze the
perpendicular correlation only, since a parallel slow-mode component is not
expected to survive but to undergo strong Landau damping. However, it is also
important to state that our simulation results cannot be unequivocal on this
issue, because at other positions and different time intervals the
anti-correlation is not always pronounced, and some cases even show a
positive correlation. These findings should be understood just as an
indication for the presence of PBSs. In this special case, the correlation
remains positive for only a few averaging steps in time.

The temperature of the treated protons stays mainly constant in this case.
Also a potential temperature anisotropy is not observed.

\subsection{Results for simulation Run B}

\begin{figure}
\includegraphics[width=\columnwidth]{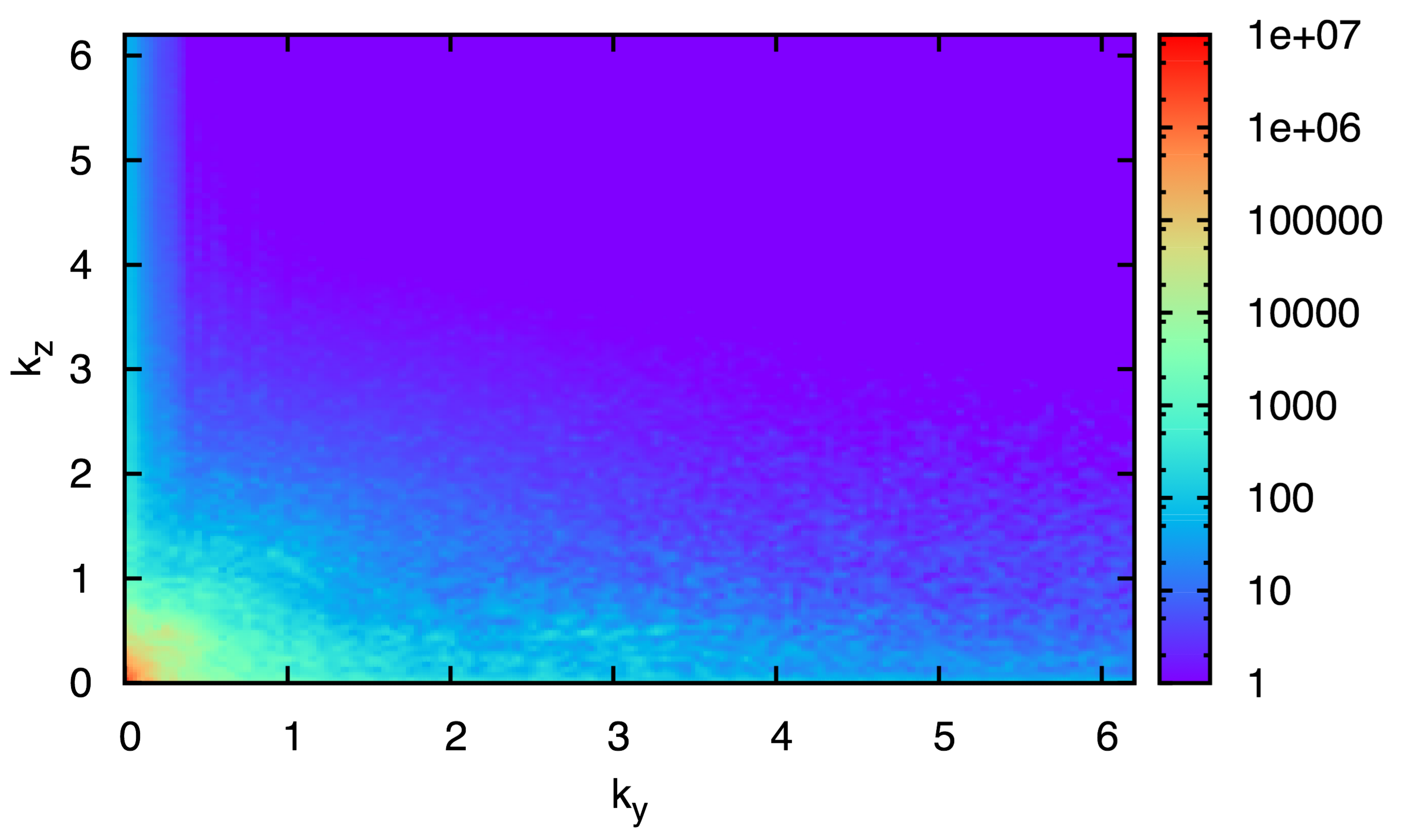}%
\caption{Two-dimensional power spectral density of magnetic field fluctuations
in arbitrary units for Run B. The background magnetic field is oriented along the $k_z$-axis.
Also in this case, the cascade is preferentially perpendicular.
\label{fig_bmhd_highamp_B_spectrum_later}}%
\end{figure}

Run B with a higher initial amplitude is in general consistent with the
previously discussed Run A. Therefore, we limit our description to the major
similarities and differences between the two runs. The magnetic field
spectrum is shown in Fig.~\ref{fig_bmhd_highamp_B_spectrum_later}.
The cascade is also preferentially oriented in the perpendicular direction.
However, a larger wavenumber range is filled isotropically around $\vec k=0$.
The compressive component is also very similar to the one in Run A (not shown
here).

\begin{figure}
\includegraphics[width=\columnwidth]{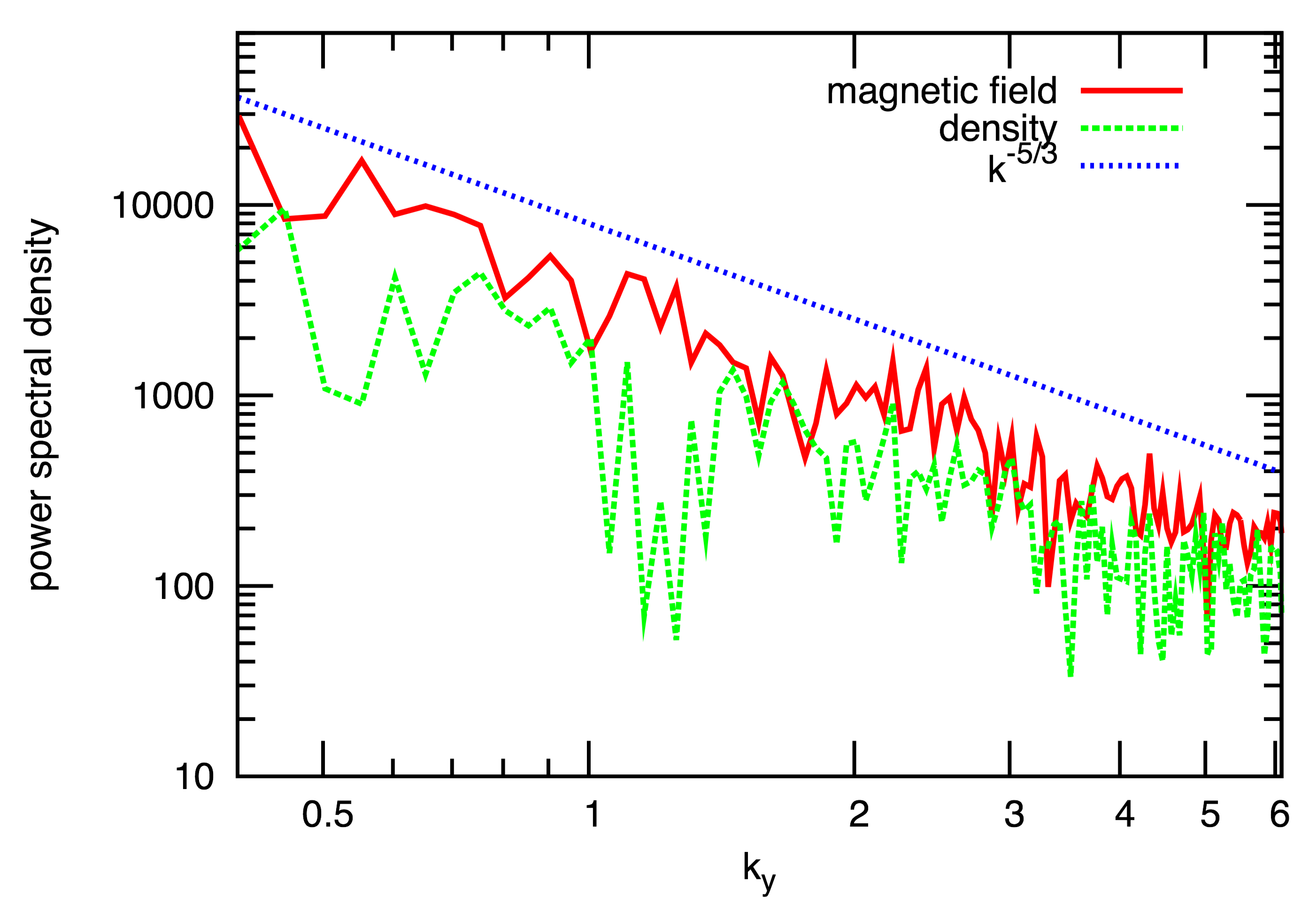}%
\caption{Power spectral density of fluctuations in the magnetic field and
density at $t=500$ along the perpendicular direction for Run B.
\label{fig_bmhd_highamp_cutspec}}%
\end{figure}

A cut along the perpendicular direction in the power spectra for the magnetic
field and density fluctuations is shown in
Fig.~\ref{fig_bmhd_highamp_cutspec}.
In this case, the spectrum follows more closely a slope with the power index of $-5/3$.

\begin{figure}
\includegraphics[width=\columnwidth]{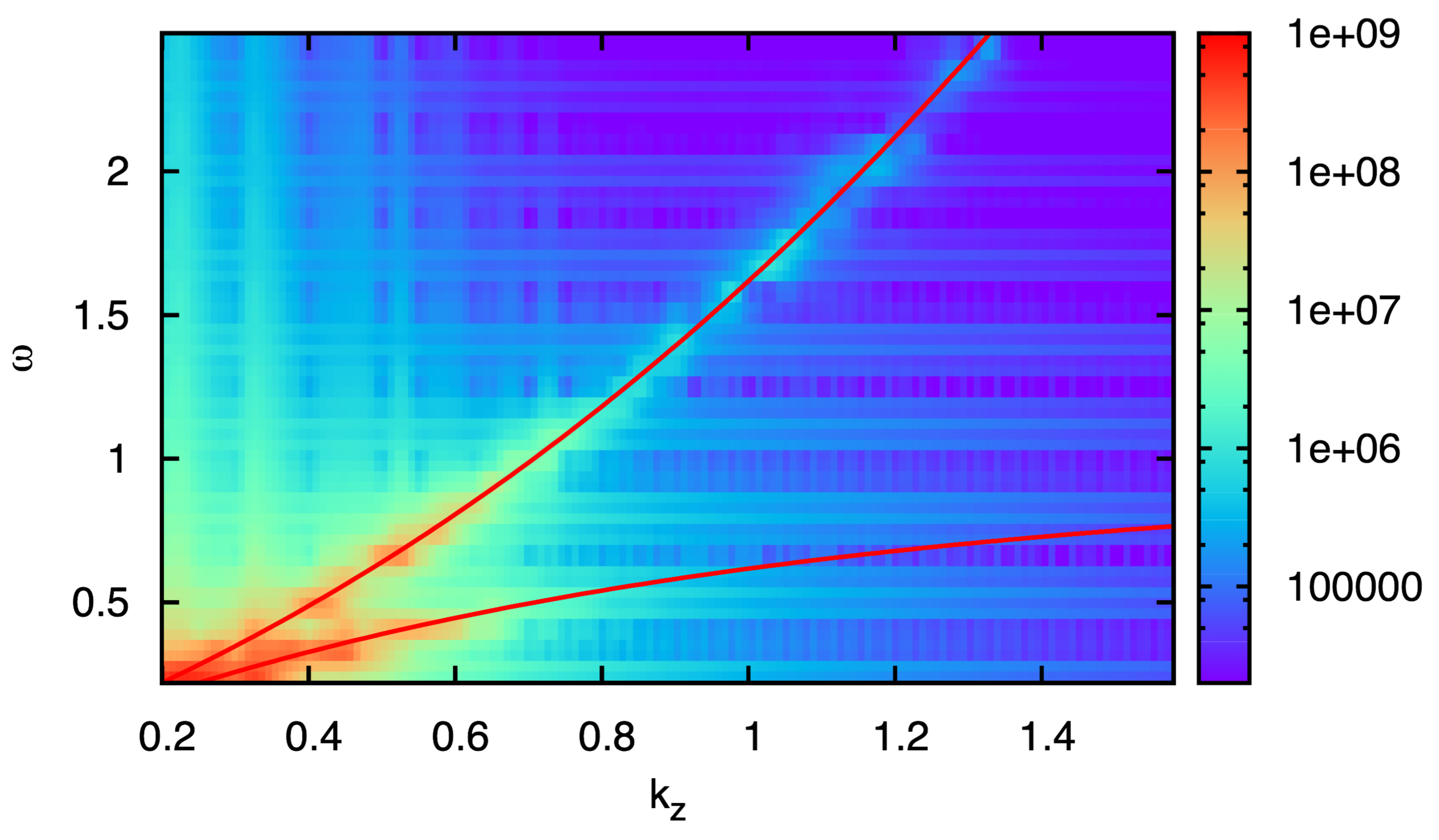}%
\caption{Dispersion relation of magnetic field fluctuations parallel to the
background magnetic field at $t\approx 500$ for Run B. The overplotted red lines
indicate again the theoretical cold-plasma dispersion relations.
\label{fig_bmhd_highamp_dispersion_B_parallel}}%
\end{figure}

The dispersion analysis of the parallel fluctuations is shown in
Fig.~\ref{fig_bmhd_highamp_dispersion_B_parallel} and reveals also the same
normal mode structure as in Run A, yet with a stronger intensity.

\begin{figure}
\includegraphics[width=\columnwidth]{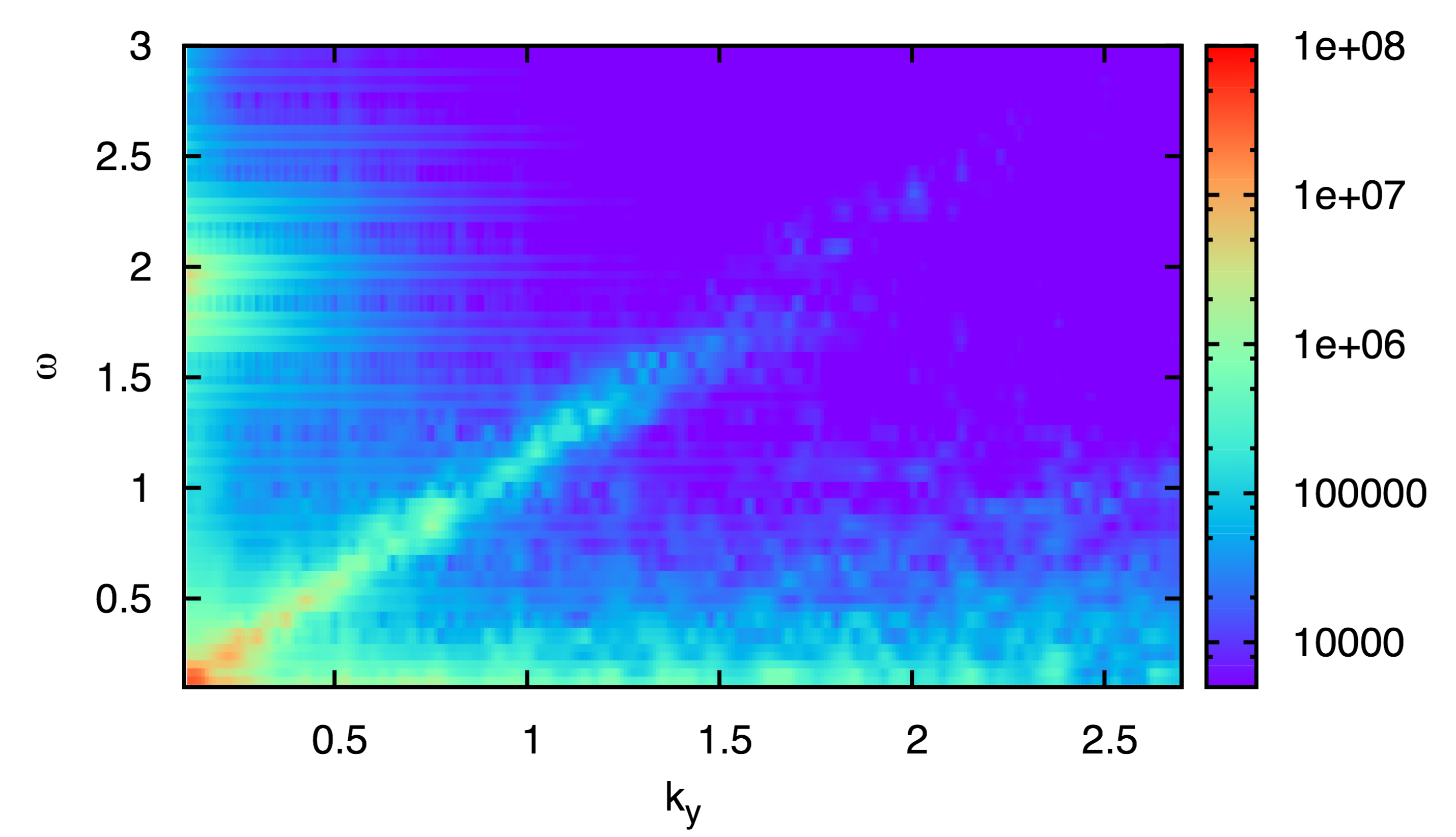}%
\caption{Dispersion relation of magnetic field fluctuations perpendicular to the
background magnetic field at $t\approx 500$ for Run B.
\label{fig_bmhd_highamp_dispersion_B_perp}}%
\end{figure}

A substantial difference occurs in the perpendicular dispersion, which is
shown in Fig.~\ref{fig_bmhd_highamp_dispersion_B_perp}.
The ion-Bernstein signatures, which have been observed in Run A, are not any
longer visible in this dispersion diagram and are also not present in the
compressive dispersion (not shown here). Instead, the perpendicular fast mode
is more pronounced.

The low-frequency part also shows an anti-correlation between $\delta P_B$
and $\delta n$ under certain conditions and, therefore, seems to represent
the same structures as observed in Run A.

\begin{figure}
\includegraphics[width=\columnwidth]{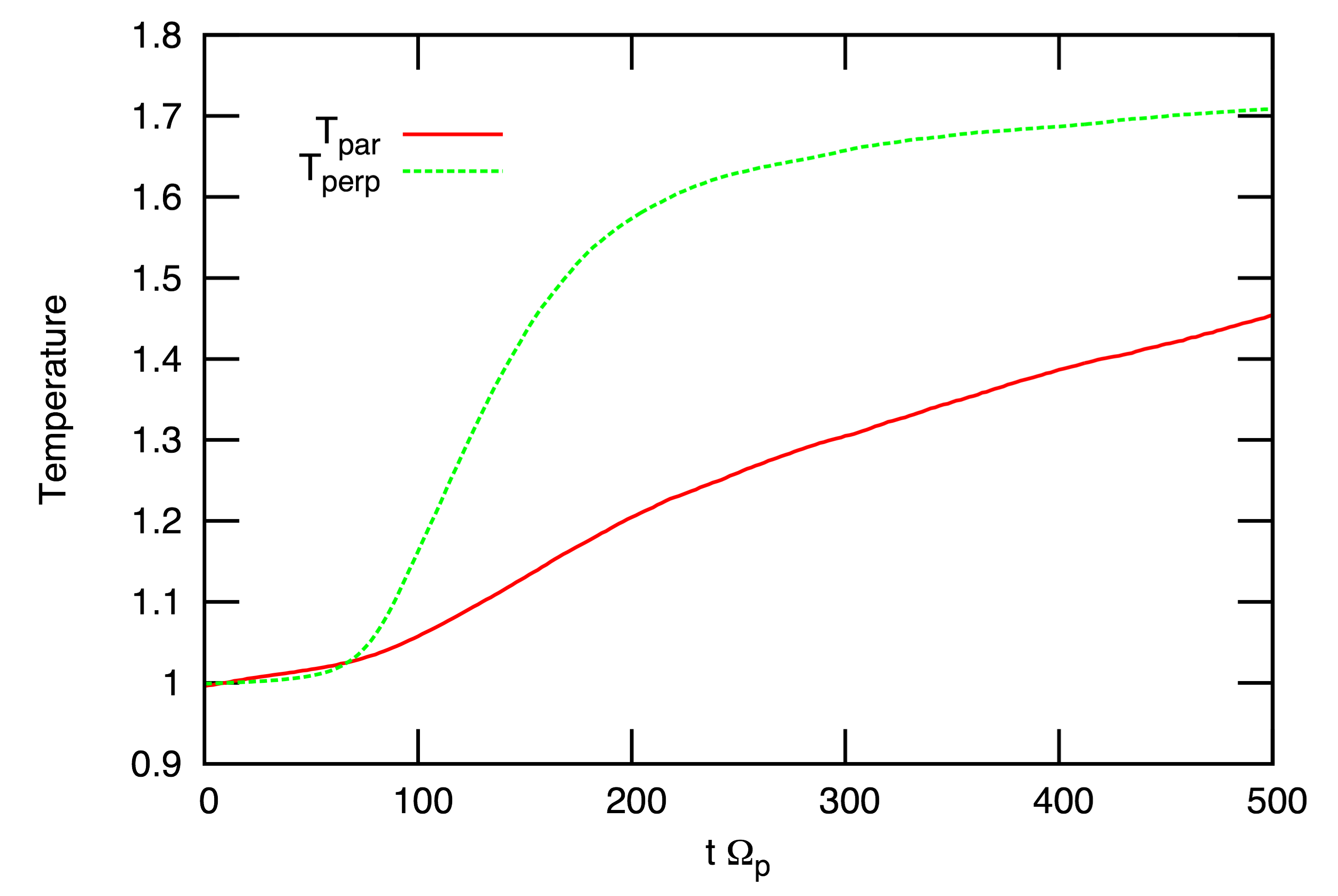}%
\caption{Temperature evolution of the protons for Run B. The temperatures versus time in the parallel and perpendicular direction with respect to the background magnetic field are shown. They are normalized to their initial values.
\label{fig_bmhd_highamp_temperature}}%
\end{figure}

While the wave power in the low-amplitude case was too low to produce
significant proton heating, this situation changed in the case of Run B. The
temperature evolution of the protons is shown in
Fig.~\ref{fig_bmhd_highamp_temperature}.
Their temperatures increase, and a temperature anisotropy with higher
perpendicular than parallel temperature with respect to the background
magnetic field is found. The perpendicular temperature raises faster than the
parallel one and reaches a saturation value of about 1.7 of the initial
perpendicular temperature.

\section{Discussion and Conclusions}

According to our numerical simulations, the turbulent energy cascades
preferentially into the direction perpendicular to the background magnetic
field, which is consistent with the recent results of other numerical
simulations\cite{macbride08,jiang09,markovskii10} and space observations
\cite{chen10,sahraoui10,narita11}. However, there is also a parallel cascade
present, and the nature of all the relevant fluctuations shall now be
discussed in dependence upon the direction of propagation.

The parallel fluctuations seem to be well described as a superposition of
normal modes. In the range above the ion-cyclotron scales they are mainly F/W
waves, as it has been suggested before by many authors
\cite{matthaeus90,stawicki01,gary08}. This finding is in agreement with
observations in the solar wind indicating that left- and right-handed normal
modes coexist until a wavelength of the order of $1/\ell_{\mathrm p}$, where
the left-handed waves undergo resonant wave-particle
interactions\cite{goldstein94} with cyclotron absorption. At higher
wavenumbers, only right-handed waves can survive the transition into the
intermediate dissipative range of solar wind turbulence \cite{he11a}. Linear
wave damping seems to dominate the dissipation compared to nonlinear damping
effects, as it has been discussed before in the context of interstellar
medium heating\cite{spangler91}.

Parallel A/IC waves are the most prominent of the left-handed waves that are
known to undergo strong ion-cyclotron damping under certain conditions
\cite{marsch06}. The observed perpendicular solar wind heating
\cite{marsch04} can, thus, be largely explained by absorption of this
left-handed normal mode, which is mostly observed in parallel propagation.
This is most probably also the explanation for the establishment of the
observed temperature anisotropy in the simulation Run B with the higher
initial amplitude. Quasi-perpendicular ion-Bernstein waves may also be able
to heat the plasma due to cyclotron-resonance effects. Therefore, also the
here identified ion-Bernstein modes could provide a heat source for the ions.
The efficiency of the dissipation of ion-Bernstein waves strongly depends on
beta and is higher for larger beta values. This effect is also discussed by
Markovskii et al. \cite{markovskii10} in the context of a cascade of F/W
waves. This may be a possible explanation for the vanishing of the
ion-Bernstein modes in Run B because the temperature increase due to the
resonant heating corresponds to a change in the plasma beta.

The reason for the flatter power index in Fig.~\ref{fig_bmhd_cutspec}
compared to the Kolmogorov scaling, which is instead found in Run B as seen
in Fig.~\ref{fig_bmhd_highamp_cutspec}, may be that in the Run A with lower
initial stirring amplitude the cascade develops more slowly, and the
turbulence is not yet fully evolved at the applied integration time.

Fast waves themselves can of course also be dissipated by ions if they
propagate obliquely \cite{marsch06}. This effect may, however, be slow when
compared to the cyclotron-resonant absorption of A/IC waves and needs to be
accumulated over a longer solar wind travel time to become significant. The
intensity of the ion-Bernstein bands is more pronounced for higher electron
betas, which is an indication for the electrostatic character of these wave
structures. A kinetic micro-instability, which is able to excite
ion-Bernstein waves in a way consistent with the wave structures observed in
magnetospheres, has recently been treated in detail with particle-in-cell
simulations\cite{liu11}.

The nature of the perpendicular low-frequency fluctuations cannot be uniquely
identified. There is evidence for the existence of pressure-balanced
structures (PBSs), which show the typical anti-correlation between $\delta
P_B$ and $\delta n$. It is observed in the solar wind that this
anti-correlation dominates on shorter time averaging\cite{tu95}. A positive
correlation dominates on longer time-scales, which is interpreted as the
indication for co-rotating interaction regions as a result of interactions
between different solar wind streams with high and low outflow speeds. The
correlation shows a typical spatial dependence on large scales in the solar
wind. A positive correlation is built up inside 0.7 to 0.8 AU, whereas the
anti-correlation is already observed closer to the Sun prevailing over a
large distance range \cite{roberts87b}. Cluster observations also show PBSs
on smaller scales than the typical low-frequency MHD range\cite{yao11}. The
origin of these structures, however, is unclear. Part of these structures may
be generated by a nonlinear cascade of the turbulence into the intermediate
wavenumber range, as it is revealed by the simulations. Other possible
non-propagating perpendicular wave structures are mirror modes or Weibel
modes.

The predominance of an anti-correlation in $\delta P_B$ and $\delta
n$ at some positions supports the interpretation of these structures in terms
of PBSs. But they are of dynamic nature and a minor component driven by the
perpendicular incompressible fluctuations which are the major component and
represent the key nonlinearly interacting degrees of freedom. Therefore,
incompressible turbulence prevails, but additionally the compressive PBS-like
correlations occur at the same time. The interpretation of such fluctuations
in terms of static PBSs would be closer to a normal-mode picture of the
modeled fluctuations but perhaps not quite adequate. However, the present
simulation results do not permit a definite conclusion on the nature of the
weakly compressive fluctuations.

In a recent model \cite{howes11}, the transition from strong to weak
turbulence at small scales is discussed. The authors assume a completely
suppressed parallel cascade for the kinetic Alfv\'en waves considered in
their model, deduced from the suppressed parallel cascade of weak turbulence
in the incompressible MHD limit. According to their model, the anisotropy in
the strongly turbulent domain is expected to follow the phenomenology of
critically balanced turbulence, which is not applicable to weak turbulence.
Our simulation shows, however, that the parallel cascade is not fully
suppressed, even not in the limit of weak turbulence. Yet, the spectral
transfer is clearly anisotropic and shows a preference for the perpendicular
cascade.

The MHD modes with low wavenumbers mainly keep their amplitude level during
their temporal evolution and stay mostly isotropic. A longer integration time
might also show a preferred direction of the cascade at lower wavenumbers, as
it was observed by other authors\cite{wicks10}. Waves with higher
wavenumbers, however, may hand over their energy more easily. This underlines
that the spectral transfer occurs also in this case locally in wavenumber
space, as it seems typical for a turbulent cascade \cite{coleman68}. This
effect is also incorporated in models concerning the coronal heating problem,
where a nonlinear local cascade is often effectively described as an
advection and diffusion process in wavenumber space\cite{zhou90,cranmer03}.

Normal modes can also be directly excited from the ubiquitous thermal
fluctuations in a plasma \cite{araneda11}. However, their amplitudes usually
stay on a very low level, and they do not increase with time at the expense
of energy drawn from lower wavenumbers, as it is observed in our simulations.
Therefore, it is reasonable to interpret the observed wave structures at
higher frequencies as being the products of processes driven from the
low-frequency side in the sense of a nonlinear mechanism, instead of being of
purely thermal origin with a typical energy content of the order $k_{\mathrm
B}T$.

If the wave power in the inertial range is much higher than assumed here,
other nonlinear couplings might play a role also in the kinetic regime. This
can possibly destroy the normal mode superposition even above $k=1$, and then
lead to a completely different picture with respect to both the spectral
transfer to higher wavenumbers and the dispersion structure in the dispersive
spectral range.

Parashar et al.\cite{parashar11} have shown that the properties of the
turbulence driver are very important for the generation and the further
evolution of plasma turbulence. The dependence on the driving frequency is
crucial in the context of AC coronal heating models, since footpoint motions
on different timescales are expected to be the main source for the
turbulence, which is then assumed to heat the coronal plasma in these models.

Finally, our simulation box is small compared to all global structures in the
solar wind, and the numerical conditions are, hence, homogeneous in the above
considerations. There are, however, recent indications that inhomogeneities
foster the thermalization of wave energy \cite{ofman11}. This effect may play
also a role for the global evolution of the solar wind.

\begin{acknowledgments}
D.~V.~received financial support from the International Max Planck Research
School (IMPRS) on Physical Processes in the Solar System and Beyond. The
calculations have been performed on the MEGWARE Woodcrest Cluster at the
Gesellschaft f\"ur wissenschaftliche Datenverarbeitung mbH G\"ottingen
(GWDG).
\end{acknowledgments}

\bibliography{beyond_mhd_paper_rev}

\end{document}